\documentclass[a4paper]{article}

\usepackage{icrc2013}

\def\hc#1{\leavevmode\hbox to \hsize{\hss #1\hss}\leavevmode}
\def\includefigure#1{\hc{\resizebox{\columnwidth}{!}{\includegraphics{#1}}}}
\def\includefigurew#1{\hc{\resizebox{0.696\textwidth}{!}{\includegraphics{#1}}}}
\paperid{1053}

\title{Progress in Monte Carlo design and optimization of the Cherenkov Telescope Array}

\shorttitle{MC design and optimization of CTA}

\authors{
   K.~Bernl{\"o}hr$^{1,2}$,         
   A.~Barnacka$^{3}$,               
   Y.~Becherini$^{4,5}$,            
   O.~Blanch Bigas$^{6}$,           
      A.~Bouvier$^{7}$,                   
   E.~Carmona$^{8,9}$,              
   P.~Colin$^{8}$,                  
   G.~Decerprit$^{10,11}$,          
   F.~Di~Pierro$^{12}$,             
   F.~Dubois$^{13}$,                
   C.~Farnier$^{14,15}$,            
   S.~Funk$^{16}$,                  
   G.~Hermann$^{1}$,                
   J.A.~Hinton$^{17}$,              
   T.B.~Humensky$^{18}$,            
     T.~Jogler$^{16}$,              
   B.~Kh\'elifi$^{4}$,              
   T.~Kihm$^{1}$,                   
   N.~Komin$^{19}$,                 
   J.-P.~Lenain$^{20}$,             
   R.~L{\'o}pez-Coto$^{6}$,         
   G.~Maier$^{10}$,                 
   D.~Mazin$^{8}$,                  
   M.C.~Medina$^{21}$,              
   A.~Moralejo$^{6}$,               
     R.~Moderski$^{3}$,             
   S.J.~Nolan$^{22}$,               
   S.~Ohm$^{17,23}$,                
   E.~de~O{\~n}a Wilhelmi$^{1}$,    
   R.D.~Parsons$^{23,1}$,           
   M.~Paz Arribas$^{9,2}$,          
   G.~Pedaletti$^{24}$,             
   S.~Pita$^{5}$,                   
   H.~Prokoph$^{10}$,                
   C.B.~Rulten$^{25}$,              
   U.~Schwanke$^{2}$,               
   M.~Shayduk$^{10}$,                
   V.~Stamatescu$^{6}$,             
   P.~Vallania$^{12}$,              
   S.~Vorobiov$^{2,10}$,             
   R.~Wischnewski$^{10}$,            
     M.~Wood$^{16}$,                
   T.~Yoshikoshi$^{26}$,            
   A.~Zech$^{25}$                   
for the CTA Consortium.
}

\afiliations{
$^{1}$ {Max-Planck-Institut f{\"u}r Kernphysik, P.O. Box 103980, 
        D-69029 Heidelberg, Germany} \\
$^{2}$ {Institut f{\"u}r Physik, Humboldt-Universit{\"a}t zu Berlin, 
        Newtonstr. 15, D-12489 Berlin, Germany} \\
$^{3}$ {Nicolaus Copernicus Astronomical Center, Polish Academy
        of Sciences, ul. Bartycka 18, 00-716 Warsaw, Poland} \\
$^{4}$ {Laboratoire Leprince-Ringuet, Ecole Polytechnique, CNRS/IN2P3,
        F-91128 Palaiseau, France} \\
$^{5}$ {Astroparticule et Cosmologie (APC), CNRS, Universit{\'e} Paris 7
        Denis Diderot, Paris Cedex 13, France} \\
$^{6}$ {IFAE, Edifici Cn., Campus UAB, E-08193 Bellaterra, Spain} \\
$^{7}$ {SCIPP, University of California, Santa Cruz, CA~95064, USA} \\
$^{8}$ {Max-Planck-Institut f{\"u}r Physik, F{\"o}hringer Ring 6, 
        D-80805 M{\"u}nchen, Germany} \\
$^{9}$ {Centro de Investigaciones Energ{\'e}ticas, Medioambientales y
        Tecnol{\'o}gicas (CIEMAT), Madrid, Spain} \\
$^{10}$ {DESY, Platanenallee 6, D-15738 Zeuthen, Germany} \\
$^{11}$ {Argonne National Laboratory, 9700 S.~Cass Avenue, Argonne, 
	    IL~60439, USA} \\
$^{12}$ {Osservatorio Astrofisico di Torino dell'Istituto Nazionale
        di Astrofisica, Corso Fiume 4, I-10133 Torino, Italy} \\
$^{13}$ {Universidad Complutense, E-28040 Madrid, Spain} \\
$^{14}$ {Department of Physics, Stockholm University, 
        AlbaNova, SE-106~91 Stockholm, Sweden} \\
$^{15}$ {The Oskar Klein Centre for Cosmoparticle Physics, 
        AlbaNova, SE-106~91 Stockholm, Sweden} \\
$^{16}$ {Kavli Institute for Particle Astrophysics and Cosmology, 
        SLAC, Stanford, CA~94025, USA} \\
$^{17}$ {Department of Physics and Astronomy, The University of Leicester,
        Leicester, LE1~7RH, United Kingdom} \\
$^{18}$ {Physics Department, Columbia University, New York, NY 10027, USA} \\
$^{19}$ {LAPP, Universit{\'e} de Savoie, CNRS/IN2P3, 
        F-74941 Annecy-le-Vieux, France} \\
$^{20}$ {LPNHE, Universit{\'e} Pierre et Marie Curie Paris 6, Universit{\'e} Denis Diderot Paris 7,
    CNRS/IN2P3, 4 Place Jussieu, F-75252 Paris Cedex 5, France} \\
$^{21}$ {CEA Saclay, DSM/IRFU, F-91191 Gif-Sur-Yvette Cedex, France} \\
$^{22}$ {University of Durham, Department of Physics, 
        South Road, Durham DH1~3LE, United Kingdom} \\
$^{23}$ {University of Leeds, School of Physics and Astronomy, 
        Leeds LS2~9JT, United Kingdom } \\
$^{24}$ {Institut de Ci{\`e}ncies de l'Espai (IEEC-CSIC), Campus UAB, Torre C5,
        E-08193 Barcelona, Spain} \\
$^{25}$ {LUTH, Observatoire de Paris, CNRS, Universit{\'e} Paris Diderot, 
        5~place Jules Janssen, F-92190 Meudon, France} \\
$^{26}$ {Institute for Cosmic Ray Research, The University of Tokyo, Kashiwa,
        Chiba 277-8582, Japan} \\
}

\email{Konrad.Bernloehr@mpi-hd.mpg.de}

\abstract{
The Cherenkov Telescope Array (CTA) will be an instrument covering a
wide energy range in very-high-energy (VHE) gamma rays. CTA will 
include several types of telescopes, in order to optimize the 
performance over the whole energy range.
Both large-scale Monte Carlo (MC) simulations of CTA super-sets
(including many different possible CTA layouts as sub-sets) and
smaller-scale simulations dedicated to individual aspects were
carried out and are on-going. We summarize results of the prior
round of large-scale simulations, show where the design has now
evolved beyond the conservative assumptions of the prior round
and present first results from the on-going new round of MC
simulations.
}


\begin{document}
\maketitle

\section{Introduction}

The Cherenkov Telescope Array (CTA) \cite{cta-www,cta2010} is planned as the 
next big step in ground-based very-high-energy (VHE) gamma-ray astronomy,
with one installation planned in the southern hemisphere and one in
the northern hemisphere. Not only will it enhance
the sensitivity by about an order of magnitude over existing instruments
but also cover a very large energy range of about four orders of magnitude,
the latter at least at the southern site.
The most cost-efficient way to achieve these goals is to build CTA
with several types of Cherenkov telescopes - a few large-size (and expensive)
telescopes (LSTs) for detecting and measuring low-energy showers, a rather
large number of mid-size telescopes (MSTs) for the core of the energy range,
plus an even larger number of small-size telescopes (SSTs) for energies in
the tens to hundreds of TeV. It may eventually be extended with
further types of telescopes -- currently foreseen are high angular resolution
MST-class telescopes of Schwarzschild-Couder type optics (SC-MST)~\cite{sc-mst}.

The evaluation of the expected performance of the different telescope designs,
of sub-arrays of equal telescopes (LSTs/MSTs/...) as well as the combined performance 
of the whole CTA instruments planned for the southern and northern hemispheres is
evaluated by the Monte-Carlo simulation method. These simulations are
using CORSIKA \cite{corsika} for the simulation of the particle showers in
the atmosphere and {\tt sim\_telarray} \cite{ber2008} for the detector
simulation. Different analysis methods have been applied to the resulting data.
See \cite{cta-mc-ApP-spec} for more details.

\section{Monte Carlo simulations}

The simulations for an instrument like CTA require substantial computing resources,
in particular for simulating enough background events (mainly proton-induced showers),
due to the excellent gamma-hadron discrimination and angular resolution of the instrument.
Apart from a small number of initial simulation sets for demonstrating that the
expected performance of CTA is not unreasonable and a large number of small-scale
simulations for optimization of the individual telescope types, the main effort
has gone and is still going into two large-scale simulation sets. 

\begin{figure}[htbp]
\includefigure{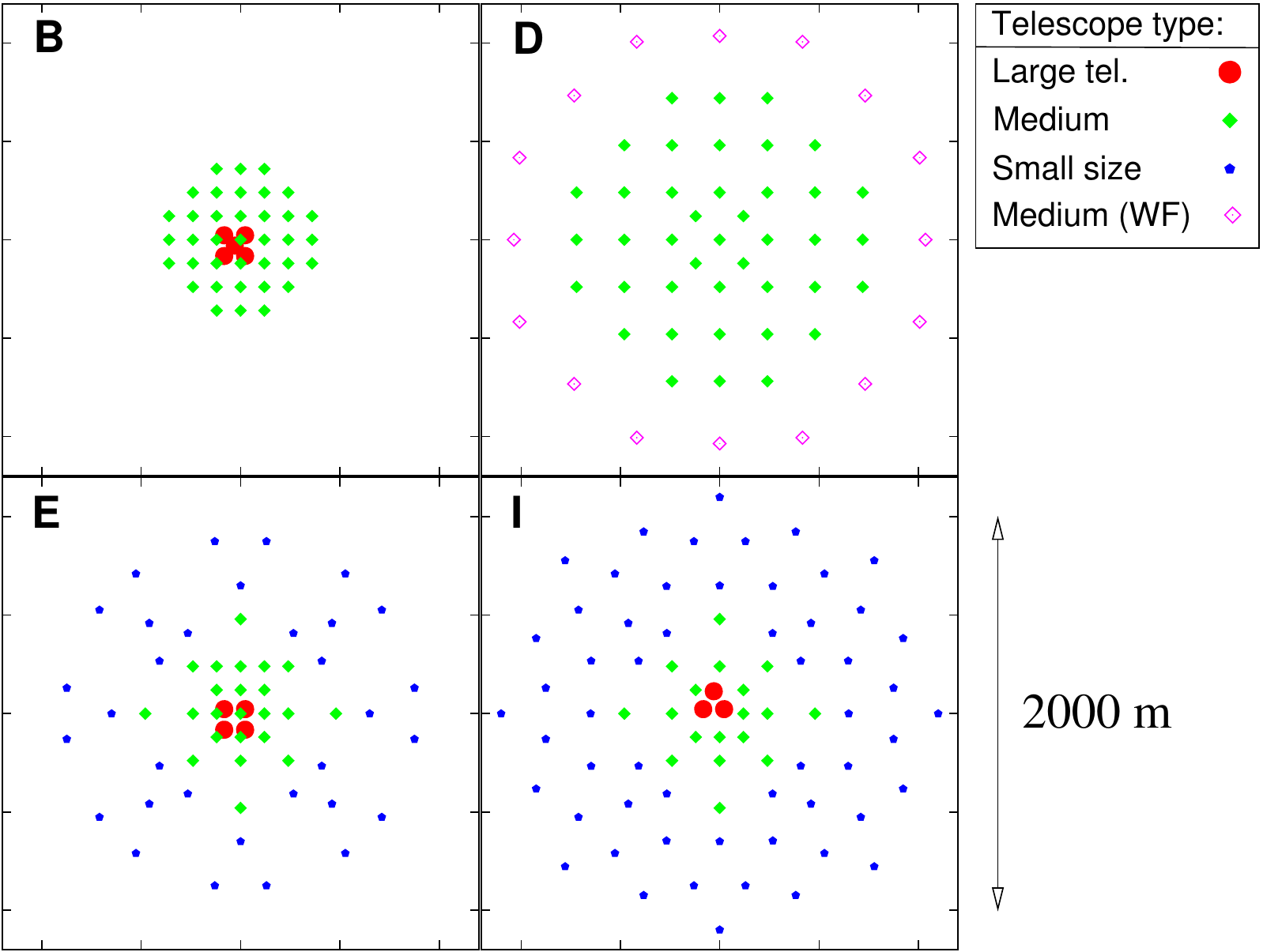}
\caption[Array I layout]{
A selection of layout candidates for a southern CTA site.
Top row: Array `B' (best low-energy performance) and Array `D'
(best high-energy performance).
Bottom row: Intermediate layouts with the best overall
physics performance, Arrays `E' and `I', the latter with
3 LSTs of 412~m$^2$ mirror area, 18 MSTs of 100~m$^2$, and
56 SSTs of 37~m$^2$.
}
\label{fig:array-I-layout}
\end{figure}

The first one, termed {\em prod-1}, was based on initial and conservative assumptions of
telescope parameters. It was carried out for hypothetical sites at altitudes of
2000~m and 3700~m, respectively. Part of these simulations were set up to correspond to
an elevated nightsky background, corresponding to partial moon light.
In all of these prod-1 simulations a total of 275 telescopes
was simulated, including five different types of telescopes. The performance parameters
as evaluated for many different subsets, each matching a given cost envelope,
were subjected to many different astrophysical test cases. These tests narrowed down the
configurations or layouts with overall best performance to a class of 
{\em intermediate layouts}, although individual astrophysical problems could be
be better studied with more compact or more widely spaced arrays.
The preferred intermediate-layout candidate, `Array I', is illustrated
in Figure~\ref{fig:array-I-layout}.
See \cite{cta-mc-ApP-spec} for the overall prod-1 layout, the assumed telescope types,
and details of evaluated subsets.

\begin{figure}[htbp]
\includefigure{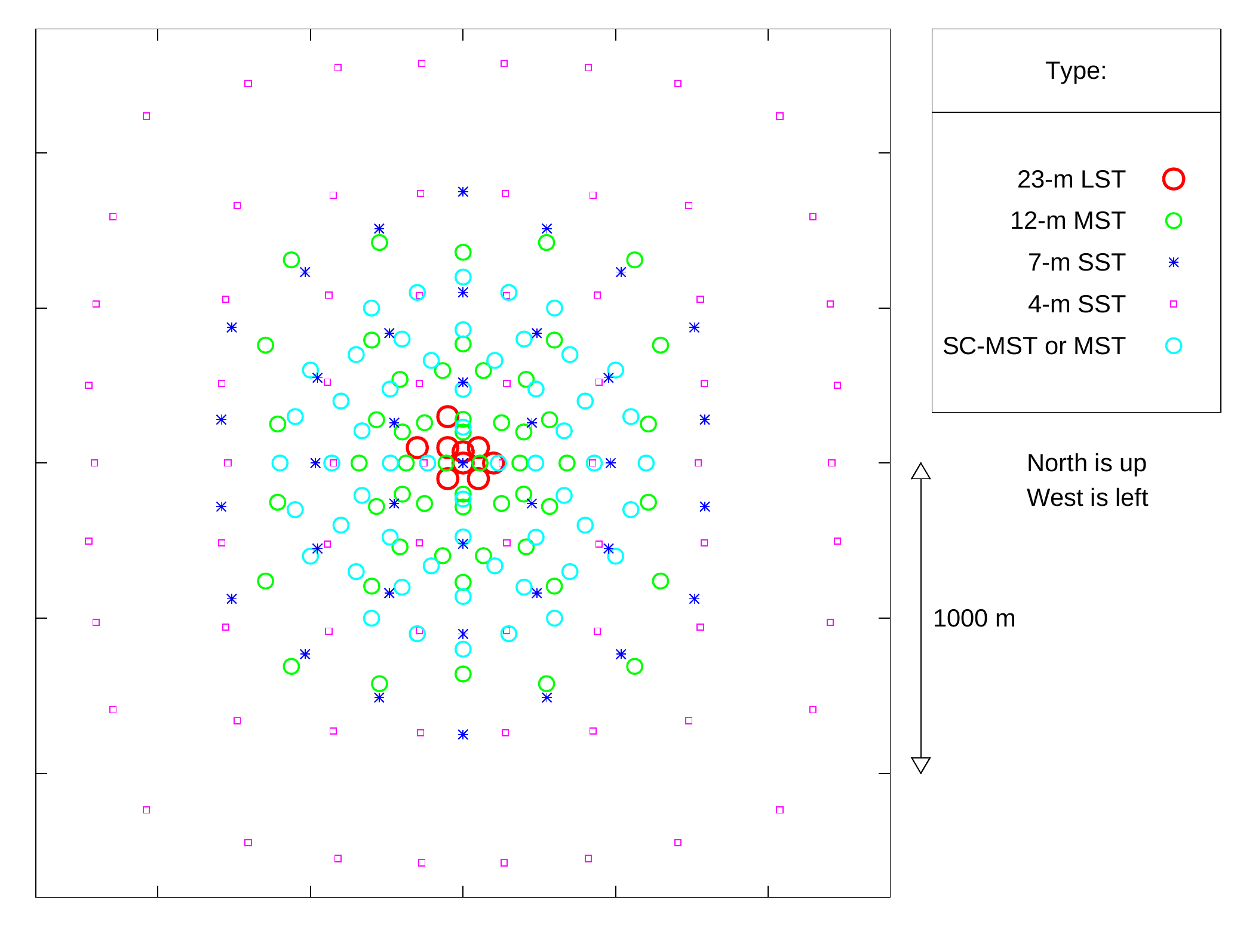}
\caption[Prod-2 layout]{
The layout of telescope positions included in the prod-2 round of large-scale
simulations for CTA design and optimization.
}
\label{fig:prod2-layout}
\end{figure}

The second round, {\em prod-2}, takes these results into consideration in the
layout of its 229 telescope positions, some of them used for more than
one type of telescope. A total of seven different types of telescopes
are included in the simulations (two different types of MSTs and four
different types of SSTs).
See Figure~\ref{fig:prod2-layout} for the overall prod-2 layout.
Telescope parameters were also adapted to current designs, including optical
design, camera design, photosensor parameters, as well as trigger and readout.
For several telescope types the simulations handle different kinds of
telescope-level triggers in parallel, such that they can be evaluated and 
compared at the level of final instrument performance -- like sensitivity.
The prod-2 simulations are currently being carried out for three different 
candidate sites at altitudes between 1600 and 3600~m.
While prod-1 only recorded one ADC sum per read-out channel, the prod-2
data includes traces (samples) of pulses in all pixels, allowing for
more advanced signal measurement methods.

\begin{figure}[htbp]
 \includefigure{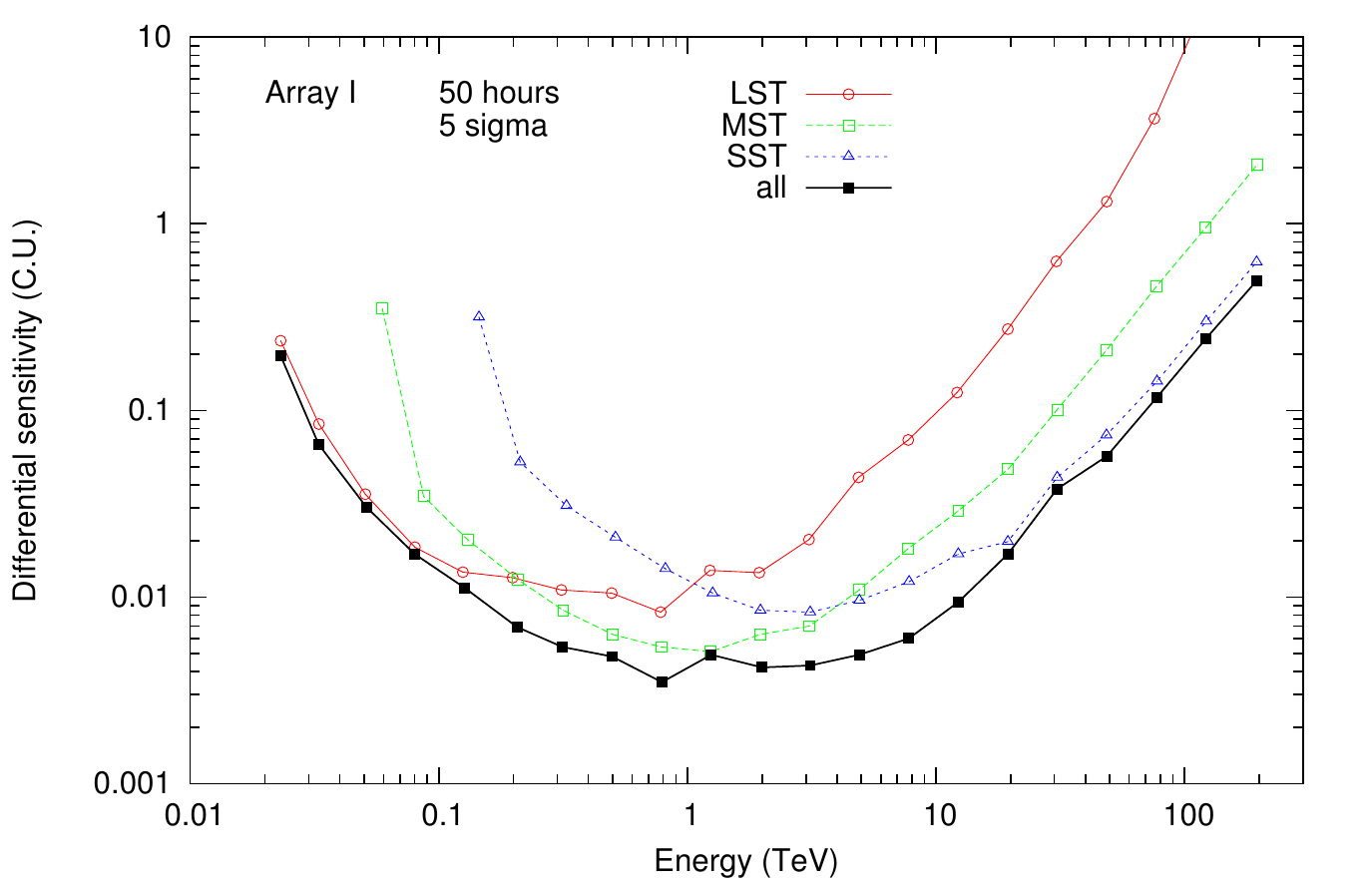}
 \caption[Sensitivity of array I and components]{
 On-axis differential point source sensitivity of one subset of prod-1 (``Array I'', 
 solid black line with filled squares) and
 its components, 3 LSTs (red, open circles), 18 MSTs (green, open squares), 
 56 SSTs (blue, open triangles) in 50 hours of observation time, 
 as derived with the baseline analysis method
 at 20$^{\circ}$ zenith angle, for a site at 2000~m altitude.
 Differential sensitivity here assumes an independent
 detection (5 sigma significance, $\ge$10 excess events, and more than 5\%  
 of the remaining background) in each energy bin.
 One Crab Unit (C.U.) here is $2.79\cdot10^{-7}$/(m$^2$ s TeV)$\times$($E$/TeV)$^{-2.57}$.}
 \label{fig:array_I_components}
\end{figure}

\section{Analysis}

\begin{figure}
    \includefigure{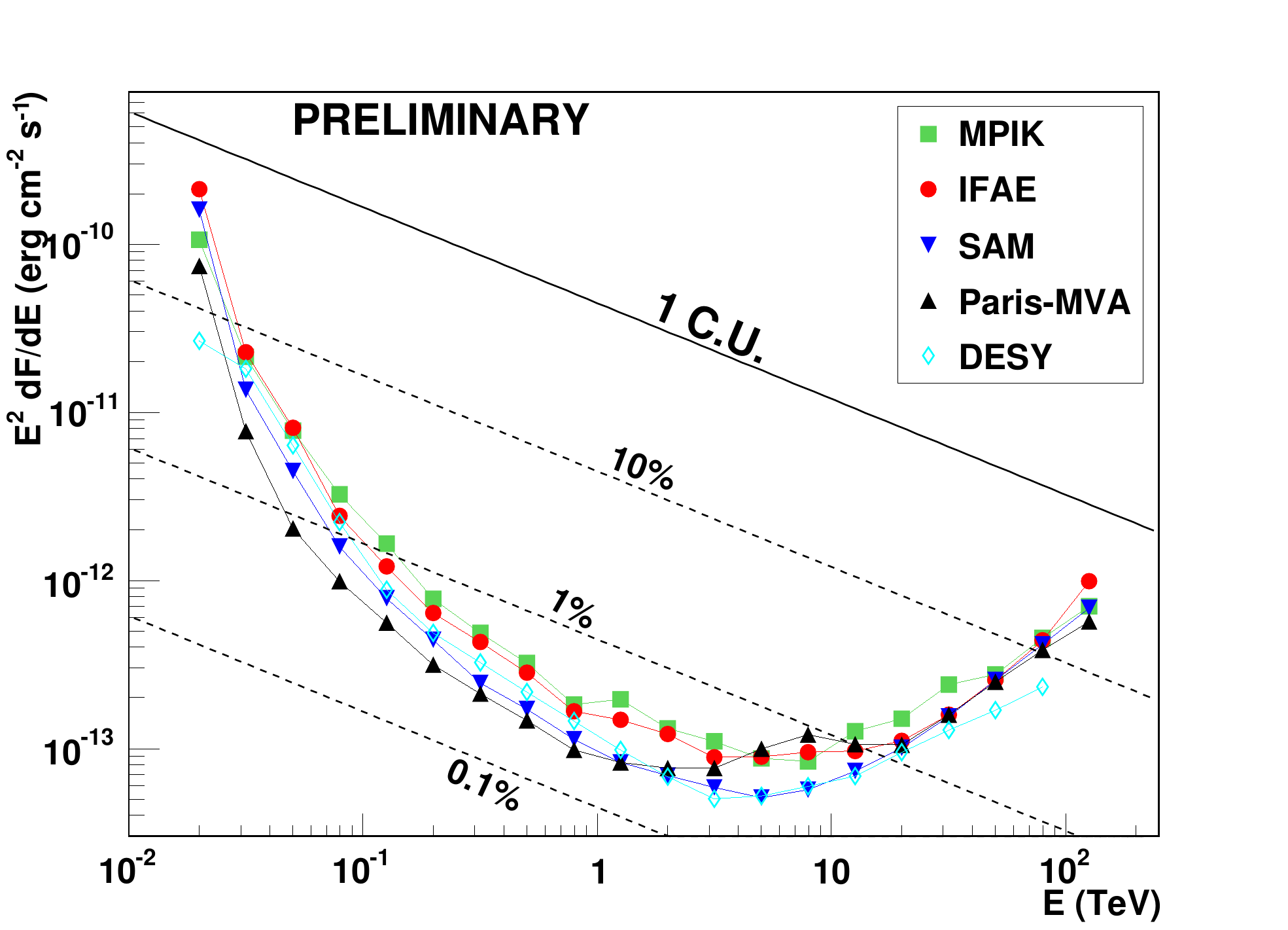}
    \caption[Differential sensitivity in alternate analyses]{
      Differential flux sensitivity of layout candidate `Array I'
      given as a function of the estimated energy,
      for the baseline/MPIK (green squares), IFAE (red circles),
      SAM (blue downward triangles) and Paris-MVA (black upward triangles) analyses
      \cite{cta-mc-ApP-spec} as well as the DESY analysis (cyan diamonds) \cite{ana-desy}.
      The Crab Unit (C.U.) flux (solid black line) 
       is shown for comparison,
      together with its $10\%$, $1\%$ and $0.1\%$ flux levels (black dashed lines).
      The differential sensitivities are optimized for an observation time of 50~h.
    }
    \label{fig:diffsens}
\end{figure}

Several sets of analysis tools \cite{cta-mc-ApP-spec} were used to process
the MC data and to evaluate the expected instrument performance.
Some of these tools were derived from 
the analysis tools of current Cherenkov telescope systems like H.E.S.S., MAGIC, 
and VERITAS, while others were developped mainly for the purpose of CTA 
MC data analysis. The baseline analysis method is basically following traditional
Hillas-parameter based stereo analysis methods, with a few additional gamma-hadron
selection cuts. Figure~\ref{fig:array_I_components} shows the expected
sensitivity of the intermediate-layout `Array I' subset of prod-1 derived with
the baseline analysis method, for 50 hours of observation time.
Some of the advanced analysis methods make use of additional
information like the time gradient along the images or image profiles, 
some apply simultaneous fits to all images. All of the advanced methods use
some machine-learning method like Neural Networks, Random Forrest, or Boosted
Decision Trees for gamma-hadron selection. As a result, the advanced methods can
achieve quite substantial improvements in sensitivity as compared to the baseline
analysis method, at least in parts of the wide CTA energy range.
A comparison of the expected sensitivity for `Array I' in different analyses is
shown in Figure~\ref{fig:diffsens}.

\section{Selected results from prod-1 simulations}

The prod-1 round of simulations demonstrated that the initial
expectations on the CTA performance were quite realistic, except perhaps
at the lowest energies where gamma-hadron selection capabilities are
limited by shower fluctuations and possible systematical errors in the
subtraction of remaining backgrounds have to be taken into account.
As Figure~\ref{fig:array_I_components} demonstrates, CTA will achieve
a high sensitivity down to energies of about 20 GeV, even with the
very conventional photo-multipliers assumed in prod-1 simulations,
with the few LSTs being responsible for the sensitivity below 100 GeV,
where the MSTs start taking over. While SSTs of the 7-m class could
have thresholds as low as 200~GeV, their wide separation prevents
high-quality data from SSTs alone below a few TeV. The sensitivity
can be expected to be dominated by the MSTs between about 200 GeV
and 4 TeV, with MSTs dominating to even higher energies when
high quality data is required for the best possible angular resolution.
Above a few TeV -- depending on the implementation -- the much larger
area covered by the SSTs (at the southern site) results in effective
detection areas growing to several square kilometers, for some
layout candidates close to 10~km$^2$.

\begin{figure*}[tbp]
    \includefigurew{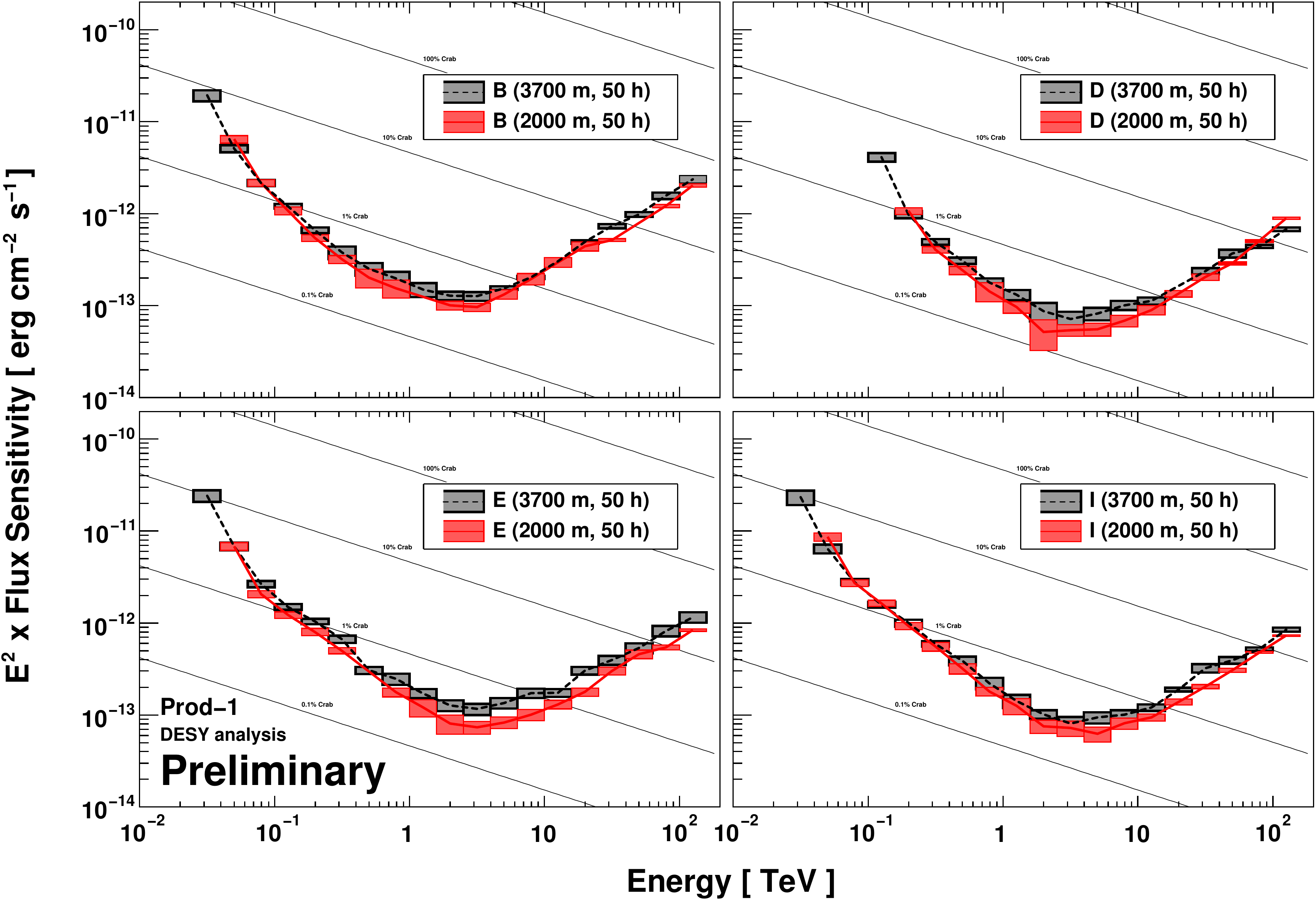}
    \caption[Differential sensitivity in alternate analyses]{
      Differential sensitivity of the four layout candidates from
      Figure \ref{fig:array-I-layout} at 2000~m altitude (red, solid line) and
      3700~m (black, dashed line), based on the DESY analysis (20$^\circ$ zenith
      angle, 50~h observation time). Apart from the different altitude, the
      simulations are for the same hypothetical site configuration.}
    \label{fig:cmp-2000-3700}
\end{figure*}

An important aspect of CTA simulations is related to site-selection
criteria, in particular the altitude of the observatory but also
the geomagnetic field. A high-altitude site has, in terms of
energy threshold, the benefit of being closer to the shower maximum,
as discussed in more detail below.
A large magnetic field, on the other hand, deflects charged particles,
spreading out the resulting Cherenkov light over a larger area
and hampering the shower reconstruction and gamma-hadron discrimination.
For a study of the combined impact of
altitude and geomagnetic field on the energy threshold of CTA
see \cite{szanecki}, based on simulations of four LSTs.
The impact of different site altitude alone on a full CTA installation is
illustrated in Figure \ref{fig:cmp-2000-3700} for the four layout
candidates shown in Figure \ref{fig:array-I-layout} and 
discussed in more detail below.

\section{Work in progress}

The prod-2 round of CTA MC production is well on the way, with simulations for the
first two of initially planned three candidate sites being close to complete and simulations
for the third candidate site ongoing (expected to be complete by mid-2013). 
The main bulk of these simulations is intended for
evaluation of the relative advantages of different site altitudes at different
energies, extending the altitude studies from prod-1 shown in Figure~\ref{fig:cmp-2000-3700}.
The lowest energies are seen to benefit from a high-altitude
site -- being closer to the shower maximum, and the Cherenkov light
less spread out as a consequence, the energy threshold will always be lower
at a high altitude. At higher energies -- already below 100 GeV -- the situation
gets more complex since at a very high altitude (above 4000 to 5000~m) more and
more particles may reach ground level, complicating the shower reconstruction
and gamma-hadron discrimination. Most of these ground-level particles appear close to
the shower axis while multi-TeV showers can be observed at larger impact parameters.
For these high energies, a high-altitude site is clearly a disadvantage since the
lateral distribution of Cherenkov light falls off more rapidly at high altitudes
(smaller detection area) and light from the shower maximum is seen at larger
angles w.r.t.\ the shower direction (large instrument field-of-view required,
with cost implications). The main task of the prod-2 round will be to find
a good compromise between the lowest possible energy threshold and the largest
possible high-energy detection area, at any given cost of a CTA installation.

Another aspect related to the CTA site selection process is the
evaluation of the performance penalty for a site with elevated nightsky
brightness. For this purpose, some of the simulations are being
reprocessed for a NSB brightness elevated by up to a factor of three.

In addition, some of the simulations include alternate instrument set-ups,
matching latest designs, in order to evaluate the relative merits to be
expected from these designs. These include three different options of
4-m class SSTs \cite{astri,gate,4m-DC} (in addition to a 7-m class SST type) as well as 
the 9-m class mid-size telescopes
with Schwarzschild-Couder dual-mirror optics (SC-MSTs)~\cite{sc-mst}.
An extension of the southern CTA site by 36 SC-MSTs
is being considered and would result in a substantial sensitivity
improvement in the key energy range between 300 GeV and 3 TeV~\cite{sc-mst2}.

A longer-term task is the continuing improvement of reconstruction,
calibration and analysis methods for the CTA observatory.
In terms of reconstruction includes the best possible measurement
of the original direction and energy of incoming gamma rays and
the discrimination between gamma rays and the background by other
particles, in an installation with several different types of
telescopes. In comparison to current instruments,
with different telescopes at different sites, the calibration
of CTA telescope systems benefits from the cross-calibration
capability between the different telescope types at the same site. 
In addition, current practices for the calibration of
the instruments can be (and will be) improved through better monitoring of
atmospheric conditions and instrument response, also checking
how well simulations correspond to measured conditions, and how
remaining deviations can be accounted for in the analysis.

\section{Conclusions}

With the previous (prod-1) and the current (prod-2) simulation rounds,
the CTA design and layout optimization can be expected to yield
quantitative results on the merits of different site altitudes,
over the complete energy regime to be covered by the Cherenkov Telescope
Array. The prod-2 round will also provide a comparison of different options for
the smaller size telescopes, in terms of expected performance of the
full instrument. Finally, the recording of traces for all pixels in
the prod-2 telescopes should help to settle the question how large
the benefit of a more advanced signal measurement (implying larger
data rates and more computing efforts) will be on the overall
CTA instrument performance.

\vspace*{0.3cm}
\footnotesize{{\bf Acknowledgment:}{We gratefully acknowledge support from the agencies and organizations 
listed in this page: \\
\url{http://www.cta-observatory.org/?q=node/22}}}

\end{document}